\documentstyle[psfig,epsf,11pt,a4,palatino]{article}
\input psfig.tex
\def\lsim{\lower.5ex\hbox{$\; \buildrel < \over \sim \;$}}
\def\gsim{\lower.5ex\hbox{$\; \buildrel > \over \sim \;$}}
\def\AHR{analogue Hawking radiation}
\def\AH{acoustic horizon}
\def\vh{\bf {{\vert_{(r=r_h)}}}}
\def\rh{\bf {(r=r_h)}}
\def\egam{$\left\{{\cal E},{\gamma}\right\}$}
\def\eker{$\left[{\cal E},\lambda,\gamma,a\right]$}
\begin{document}
\baselineskip = 14pt
\begin{center}
{
{\LARGE\bf Transonic Black Hole Accretion as Analogue System}\\[0.25cm]
{\it Tapas Kumar Das\\
N. Copernicus Astronomical Centre, Bartycka 18, Warsaw 00-716, Poland\\
tapas@camk.edu.pl}}\\[0.5cm]
\end{center}
\begin{abstract}
\noindent
Classical black hole analogues (alternatively, the analogue systems) are fluid dynamical 
analogue of general relativistic black holes. Such analogue effects may be  observed when
acoustic perturbations (sound waves) propagate through a classical dissipation-less 
tran-sonic fluid. The acoustic horizon, which resembles the actual black hole event horizon 
in many ways,
may be generated at the transonic point in the fluid flow. Acoustic horizon emits
quasi thermal phonon spectra, which is analogous to the actual Hawking radiation,
and possesses the temperature
referred as the
analogue Hawking temperature, or simply, the  analogue temperature.

Transonic accretion onto astrophysical black holes is a very interesting example of
classical analogue system found naturally in the Universe. An accreting black holes system 
as a classical analogue is unique in the sense that only for such a system,
both kind of horizons, the electromagnetic and the acoustic (generated due to
transonicity of accreting fluid) are simultaneously present in the same system. 
Hence an accreting astrophysical black hole is the ideal-most candidate to theoretically 
study and to compare the properties of these two different kind of horizons. Also such 
system is unique in the aspect that general relativistic 
spherical accretion onto the Schwarzschild black hole represents the only 
classical analogue system found in the nature so far,
 where the analogue Hawking temperature
may be higher than the actual Hawking temperature. 
\end{abstract}
\section{Black Holes}
\noindent
Black holes are vacuum solutions of Einstein's field equations in general
relativity. Classically, these objects are conceived as singularities in space
time, censored from the rest of the Universe by mathematically defined one way
surfaces, the event horizons. The space time metric defining the vacuum exterior
of a classical black hole, and the black hole itself, is characterized by only
three parameters, the mass of the black hole $M_{BH}$, the rotation (spin) $J$
and charge $q$. For $J=q=0$, one obtains a Schwarzschild black hole, and for $q=0$ 
one obtains a Kerr black hole. These two kind of black holes are important in
astrophysics. In astrophysics, black holes are the end point of gravitational
collapse of massive celestial objects. Astrophysical black holes may be broadly
classified into two categories, the stellar mass ($M_{BH}{\sim}$ a few 
$M_{\odot}$, where $M_{\odot}$ is the mass of the Sun), 
and super massive ($M_{BH}{\ge}{10^6}M_{\odot}$) 
black holes (SMBH).
While the birth history of the stellar mass black holes is theoretically known
with almost absolute certainty (they are the endpoint of the gravitational
collapse of massive stars), the formation scenario of the supermassive black
hole is not unanimously understood. 
A SMBH may form through the
monolithic collapse of early proto-spheroid gaseous mass originated at the time
of galaxy formation. Or a number of stellar/intermediate mass 
black holes may merge to form it. Also the
runaway growth of a seed black hole by accretion in a specially favoured high-density
environment may lead to the formation of SMBH. However, it is yet to be well understood
exactly which of the above processes routes towards the SMBH formation, 
see, e.g. Rees 2002 for a comprehensive
review on the formation and evolution of SMBH.
Both kind of astrophysical black holes,
the stellar mass and SMBH,
however, accrete matter from the surrounding. Depending on the intrinsic
angular momentum content of accreting material, either spherically symmetric
(for zero angular momentum flow), or
axisymmetric (for flow with non-zero finite angular momentum)
flow geometry is invoked to study an accreting black hole system
(Frank, King \& Raine 1992).
\section{Black hole thermodynamics}
\noindent
Within purely classical framework, black holes in any diffeomorphism covariant
theory of gravity (where the field equations directly follow from the
diffeomorphism covariant Lagrangian) and in general relativity, mathematically
resembles some aspects of classical thermodynamic systems
(Wald 1984, 1994, 2001, Kiefer 1998, Brown 1995 and references therein). 
In early seventies, a
series of influential works 
(Bekenstein 1972, 1972a, 1973, 1975, Israel 1976,
Bardeen, Carter \& Hawking 1977, see also Bekenstein 1980 for a review)
revealed the idea that classical black
holes in general relativity, obey certain laws which bear remarkable analogy to the
ordinary laws of classical thermodynamics. Such analogy between black hole
mechanics and ordinary thermodynamics ('the generalized Second Law', as it is
customarily called) leads to the idea of the `surface gravity' of black
hole,\footnote {The surface gravity may be defined as the acceleration measured
by red-shifts of light rays passing close to the horizon (Helfer
2003)} $\kappa$, which can be obtained by computing the norm of the gradient of
the norms of the Killing fields evaluated at the stationary black hole
horizon, and is constant on the horizon (analogous to the constancy of
temperature T on a body in thermal equilibrium - the Zeroth Law of classical
thermodynamics). Also, $\kappa$=0 can not be reached by performing finite number of
operations (analogous to the 'weak version' of the third law of classical
thermodynamics where temperature of a system cannot be made to reach at absolute
zero, see discussions at Kiefer 1998). It was found by analogy via black
hole uniqueness theorem (see. e.g. Heusler 1996 and references therein) 
that the role of entropy in
classical thermodynamic system is played by a constant multiple of the surface
area of a classical black hole. 
\section{Hawking Radiation}
\noindent
The resemblance between the laws of ordinary
thermodynamics to those of black hole mechanics were, however, initially
regarded as purely formal. This is because, the physical temperature of a black
hole is absolute zero (see, e.g. Wald 2001). Hence physical relationship between
the surface gravity of the black hole, and the temperature of a classical 
thermodynamic system can not be conceived. This further
indicates that a classical black hole can never radiate. However, introduction
of quantum effects might bring a radical change to the situation. In an epoch
making paper published in 1975,
Hawking (1975) used quantum field theoretic
calculation on curved spacetime to show that the physical
temperature and entropy of black hole {\it does} have finite non-zero value
(see Page 2004 for an excellent review of black hole thermodynamics and 
Hawking radiation). A
classical space time describing gravitational collapse leading to the
Schwarzschild black
hole was assumed to be the dynamical back ground, and a linear quantum field,
initially in it's vacuum state prior to the collapse, was considered to propagate
against this background. The vacuum expectation value of the energy momentum
tensor of this field turned out to be negative near the horizon. This phenomenon
leads to the flux of negative energy into the hole.  Such negative energy flux would
decrease the mass of the black hole and would lead to the fact that the quantum
state of the outgoing mode of the field would contain particles. \footnote { For
a lucid description of the physical interpretation of Hawking radiation,
see, e.g., Wald 1994, Kiefer 1998, Helfer 2003.}. The expected number of such
particles would correspond to radiation from a perfect black body of finite size.
Hence the spectrum of such radiation is thermal in nature, 
and the temperature of such
radiation, the Hawking temperature $T_H$ from a Schwarzschild black 
hole, can be computed as 
$$
T_H=\frac{{\hbar}c^3}{8{\pi}k_bGM_{BH}}
\eqno{(1)}
$$
where c and $k_b$ are the velocity of light in vacuum and the Boltzmann's
constant, respectively.
The semi classical description for
Hawking radiation treats the gravitational field classically and the
quantized radiation field satisfies the d'Alembert equation. At any time, black
hole evaporation is an adiabatic process if the residual mass of the hole at
that time is larger than the Planck mass.
\section {Towards an analogy of Hawking effect: The motivation}
\noindent
Substituting the values of the fundamental constant in eq. (1), one can rewrite
$T_H$ for a Schwarzschild black hole as :
$$
T_H ~{\sim}~ 6.2{\times}10^{-8}
\left(\frac{M_{\odot}}{M_{BH}}\right) {^oK}
\eqno{(2)}
$$
It is evident from the above equation that for one solar mass black hole, the
value of the Hawking temperature would be too small to be experimentally
detected. A rough estimate shows that $T_{H}$ for stellar mass black holes would
be around $10^{7}$ times colder than the cosmic microwave background radiation.
The situation for super massive black hole will be much more worse, as 
$T_H {\propto} \frac {1}{M_{BH}}$. Hence $T_{H}$ would be a measurable quantity
for primordial black holes with very small size and mass, if such black holes
really exist, and if instruments can be fabricated to detect them. The lower
bound of mass for such black holes may be estimated analytically. The time-scale
${\cal T}$ over which the mass of the black hole changes significantly due to
the Hawking's process may be obtained as (Helfer 2003)
$$
{\cal T}{\sim}\left(\frac{M_{BH}}{M_{\odot}}\right)^3 10^{65}
\eqno {(3)}  
$$
As the above time scale is a measure of the lifetime of the hole itself, the
lower bound for a primordial hole may be obtained by setting $ {\cal T}$ equal
to the present age of the Universe. Hence the lower bound
for the mass of the primordial black holes
comes out to be around $10^{15}$ gm. The size of such a black hole
would be of the order of $10^{-13}$ cm and the corresponding $T_{H}$ would be
about
$10^{11}{^oK}$, which is comparable, as we will see in \S  10, with the macroscopic 
fluid temperature of the freely falling matter (spherically symmetric accretion) onto
an one solar mass isolated Schwarzschild black hole. However, present day
instrumental technique is far from efficient to detect these primordial black
holes with such an extremely small dimension, if such holes exist at all in
first place. Hence, the observational manifestation of Hawking radiation seems
to be practically impossible. \\
\noindent
On the other hand, due to the infinite redshift caused by the event horizon, the
initial configuration of the emergent Hawking Quanta is supposed to possess
trans-Planckian frequencies and it's wave lengths are beyond the Planck scale;
thus low energy effective theories cannot self consistently deal with the
Hawking radiation (Parentani 2002). Also, the nature of the fundamental degrees
of freedom and the physics of such ultra short distance is yet to be well
understood. Hence the fundamental issues like the statistical meaning of the
black hole entropy, or the exact physical origin of the out going mode of the
quantum field, remains unresolved (Wald 2001).\\
\noindent
Perhaps the above mentioned issues served as the principal motivations to launch
a theory, analogous to the Hawking's one, whose effects would be possible
to comprehend through relatively more perceivable physical systems.
The theory of {\AHR}  opens up the possibility to experimentally verify some
basic features of black hole physics by creating the sonic horizons in the
laboratory. A number of works have been carried out to formulate the
condensed matter or optical analogs of event horizons
\footnote{Literatures on study of analogue systems in condensed matter or optics 
are quite large in numbers. Condensed matter or optical analogue systems
deserve the right to be discussed as separate review articles on its own. In
this article, we, by no means, 
are able to provide the complete list of references for theoretical or
experimental works on such systems. However, to have an idea on 
the analogue effects in condensed matter or optical 
systems, readers are refereed to the book by Novello, Visser \& Volovik 2002
for review, and to some of the representative papers like
Leonhard 2002, 2003, Garay, Anglin, Cirac, \& Zoller 2000,
2001, Jacobson \& Volovik 1998, Volovik 1999, 2000, 2001, 
Brevik \& Halnes 2002,  Sch$\ddot{\rm u}$tzhold, G$\ddot{\rm u}$nter \&
Gerhard 2002, Sch$\ddot{\rm u}$tzhold \& Unruh 2002, de Lorenci, Klippert \& Obukhov
2003, Reznik 2000, Novello, Perez Bergliaffa, Salim, DeLorenci \& Klippert 2003.}
It is also
expected that {\AHR}  may find important uses in the fields of 
FRW cosmology (Barcelo, Liberati \& Visser 2003), inflationary
models, quantum gravity and sub-Planckian models of string theory
(Parentani 2002). \\
\noindent
For space limitation, in this article, we will, however, mainly describe the formalism behind the 
{\it classical} analogue systems.
By `classical analogue systems' we refer to the examples where 
the analogue effects are studied in classical fluids, and not in 
quantum fluids.
In the following section, we narrate the basic features of a classical analogue
system. Hereafter, 
we shall use $T_{{AH}}$
to denote the analogue Hawking temperature, and $T_H$ to denote the
the actual Hawking temperature as defined in eq. (1).
We shall also use the words `analogue', `acoustic' and `sonic' synonymously
in describing the horizons or black holes. Also the phrases `analogue Hawking 
radiation/effect/temperature' should be taken as identical in meaning with the phrase
`analogue radiation/effect/temperature'. A system manifesting the 
effects of analogue radiation, will be termed as analogue system.
\section{Theory of analogue radiation}
\noindent
In a pioneering work, Unruh (1981) showed that a classical system,
relatively  more clearly perceivable than a quantum black hole system, does
exist, which resembles the black hole as far as the quantum thermal radiation is
concerned. The behaviour of a linear quantum field in a classical gravitational 
field was simulated by the propagation of acoustic disturbance in a convergent
fluid flow. In such a system, it is possible to study the effect of the reaction
of the quantum field on it's own mode of propagation and to contemplate the
experimental investigation of the thermal emission mechanism. If one considers
the equation of motion for a transonic barotropic irrotational fluid, it can be
shown (Unruh 1981) that the scaler field representing the acoustic perturbation
(i.e, the propagation of sound wave) satisfies a differential equation which is
analogous to the equation of a massless scaler field propagating in a metric.
Such a metric
closely resembles the Schwarzschild metric near the horizon. Thus acoustic
propagation through a supersonic fluid forms an analogue of event horizon, as
 the 'acoustic horizon' at the transonic point. The behaviour of the normal modes near
the acoustic horizon indicates that the acoustic wave with a quasi-thermal
spectrum will be emitted from the acoustic horizon and the temperature of such
acoustic emission may be calculated as:
$$ T_{AH}=\frac{\hbar}{4{\pi}k_b}\left[\frac{1}{a_s}\frac{{\partial}{u_{\perp}^2}}{\partial{n}}
\right]_{\rm acoustic~horizon}
\eqno {(4)}
$$
Where $a_s$ is the sound speed,  $u_{\perp}$ is the component of the
dynamical flow velocity normal to the acoustic horizon, and
$ \frac{\partial}{{\partial}n}$
represents the normal derivative.
Equation (4) has clear resemblance with eq. (1) and hence 
$T_{AH}$ is designated as analogue Hawking Temperature and such quasi-thermal
radiation form acoustic (analogue) black hole is known as the Analogue Hawking
radiation. Note that the sound speed in Unruh's original
treatment (the above equation) was assumed to be constant in space,
i.e., equation of state used was isothermal. \\ 
\noindent
Unruh's work was followed by other important papers (Jacobson 1991,
Unruh, 1995, Visser, 1998, Jacobson 1999, Bili$\acute{c}$ 1999).
A generalized treatment for classical
analogue radiation for 
Newtonian fluid may be obtained in Visser (1998). Visser (1998) showed that
for barotropic, inviscid, hydrodynamic fluid, the equation of motion for the velocity
potential $\psi$ describing an acoustic disturbance is identical to the 
d'Alembertian equation of motion for a minimally coupled massless scaler field
propagating in a (3+1) Lorenzian geometry as:
$$
\Delta \psi \equiv
{1\over\sqrt{-g}}
\partial_\mu
\left( \sqrt{-g} \; g^{\mu\nu} \; \partial_\nu \psi \right) = 0
\eqno{(5)}
$$
The acoustic propagation is described by the following metric, which is
algebraically dependent on the density and dynamical as well as sonic velocity of the flow
:
$$
G_{\mu\nu}(t,\vec x)
\equiv {\rho\over c}
\left[ \matrix{-(a_s^2-u^2)&\vdots&-{\vec u}\cr
               \cdots\cdots\cdots\cdots&\cdot&\cdots\cdots\cr
               -{\vec u}&\vdots& I\cr } \right]
\eqno{(6)}
$$
The above acoustic metric for a point sink, is conformally related to the 
Painlev\'e-Gullstrand-Lema\^\i{}tre  form of Scwarzschild metric. The conformal factor
may be neglected in connection to the calculation of the analogue Hawking
temperature. The surface gravity $\kappa$ for above mentioned acoustic metric
can be calculated as:
$$
\kappa=\frac{1}{2}\frac{{\partial}\left(a_s^2-u^2_{\perp}\right)}{{\partial}n}
\eqno{(7)}
$$
From eq (7), it is straight forward to
calculate the expression for analogue temperature as
$$
T_{AH}=\frac{{\hbar}}{4{\pi}k_b}
\left[\frac{1}{a_s}\frac{\partial}{\partial{n}}
\left(a_s^2-u_{\perp}^2\right)\right]_{\rm acoustics~ horizon}
\eqno {(8)}
$$
The above equation shows that one needs to know the exact location of the 
acoustic horizon, the exact values of the dynamical and sound velocities, 
and their space gradients on the acoustic horizon; to determine the 
analogue Hawking temperature of a classical analogue system.
Note that the only difference between eq (8) and eq. (4)
is that, in Unruh's original expression, positional constancy of sound speed
was assumed, whereas such assumption has not been made in eq. (8). \\
\noindent
For analogue systems discussed above, the fluid particles are coupled to
the {\it flat} metric of Mankowski's space (because the governing equation for fluid
dynamics in the above treatment is completely Newtonian), whereas the sound wave
propagating through the non-relativistic fluid is coupled to the {\it curved} 
pseudo-Riemannian metric. Phonons (quanta of acoustic perturbations) are the
null geodesics, which generate the null surface, i.e., the acoustic horizon. 
Introduction of viscosity may destroy the Lorenzian invariance and hence
the acoustic analogue is best observed in a vorticity free completely
dissipationless fluid. That is why, the 
Fermi superfluids and the Bose-Einstein condensates are ideal 
to simulate the analogue effects. 
\noindent
The most important issue emerging out of the above
discussions is the following:\\
Even if the governing equation for fluid flow is completely non-relativistic
(Newtonian), the acoustic fluctuations embedded into it are described by a curved
pseudo-Riemannian geometry. This information is useful to portray the immense
importance of the study of the acoustic black holes, i.e. the black hole
analogues, or simply, the analogue systems. \\
\noindent
In summary, analogue (acoustic)
black holes (or systems) are fluid-dynamic analogue of general relativistic black
holes. Analogue black holes possess analogue (acoustic) event horizons at local
transonic points. Analogue black holes (alternatively, analogue systems) emit
analogue Hawking radiation, the temperature of which is termed as analogue
Hawking temperature, and it's expression is formulated using Newtonian
description of fluid flow. Black hole analogues are important to study because
it may be possible to create them experimentally in laboratories to study some
properties of the black hole event horizon, and to study the experimental
manifestation of Hawking radiation. In the subsequent sections, we will 
describe how accretion processes onto astrophysical black holes 
may constitute a concrete example of analogue system naturally found in the
Universe. We will also demonstrate that accreting black holes are 
unique as analogue systems because they are the {\it only} system found in nature
so far, where the analogue temperature may exceed the actual Hawking temperature.
\section{Transonic black hole accretion in astrophysics}
\noindent
Gravitational capture of surrounding fluid by massive
astrophysical objects is known as accretion. There remains a major difference between black hole
accretion and accretion onto other cosmic objects including neutron stars and
white dwarfs. For celestial bodies other than black holes, infall of matter terminates
either by a direct collision with the hard surface of the accretor or with the outer boundary of the 
magneto-sphere, resulting the luminosity through energy release 
from the surface. Whereas for black hole accretion, matter ultimately 
dives through the event horizon from where radiation is prohibited to escape according 
to the rule of classical general relativity and the  emergence of luminosity occurs
on the way towards the black hole event horizon. The efficiency
of accretion process may be thought as a measure of the
fractional conversion of gravitational binding energy of matter
to the emergent radiation and is considerably high for black
hole accretion  compared to accretion onto any other
astrophysical objects. Hence accretion onto classical astrophysical black holes
has
been recognized as a fundamental phenomena of increasing
importance in relativistic and high energy astrophysics.
The
extraction of gravitational energy from the black hole accretion is believed to
power the energy generation mechanism of
X-ray  binaries and of the most luminous objects of the
Universe, the Quasars and active galactic nuclei (Frank, King
\& Raine 1992). The
black hole accretion is, thus, the most appealing way through which the
all pervading power of gravity is explicitly manifested.\\
\noindent
If the instantaneous dynamical velocity and local acoustic velocity
 of the accreting fluid, moving along a space curve parameterized by $r$, are
$u(r)$ and $a_s(r)$ respectively, then the local Mach number $M(r)$ of the
 fluid can be defined as $M(r)=\frac{u(r)}{a_s(r)}$.
The flow will be locally
 subsonic or supersonic according to $M(r) < 1$ or $ >1$, i.e., according to
 $u(r)<a_s(r)$ or $u(r)>a_s(r)$. The flow is transonic if at any moment
 it crosses $M=1$. This happens when a subsonic to supersonic or supersonic to
 subsonic transition takes place either continuously or discontinuously.
The point(s) where such crossing
takes place continuously is (are) called sonic point(s),
 and where such crossing takes place discontinuously are called shocks
 or discontinuities.
In order to
satisfy the inner boundary conditions imposed by the
event horizon, accretion onto black holes
exhibit transonic properties in general. \\
\noindent
Investigation of accretion processes onto celestial objects
was initiated by
Hoyle \& Littleton (1939) by computing the rate at which
pressure-less matter would be captured by a moving star. Subsequently,
theory of
stationary, spherically symmetric and transonic hydrodynamic accretion of
adiabatic fluid onto a gravitating astrophysical object at rest was
formulated in a seminal paper by Bondi (1952) using purely Newtonian potential
and by including the pressure effect of the accreting material.
Later
on, Michel (1972) discussed fully general relativistic polytropic accretion on
to a Schwarzschild black hole by formulating the governing equations for steady
spherical flow of perfect fluid in Schwarzschild metric. Following
Michel's relativistic generalization of Bondi's treatment,
Begelman (1978) and Moncrief (1980)
discussed some aspects of the critical points of the
flow for such an accretion.
Spherical accretion and wind in general relativity have also been considered
using equations of state other than the polytropic one and
by incorporating various radiative processes
(Shapiro, 1973a,b, Blumenthal \& Mathews 1976,
Brinkmann 1980).
Malec (1999) provided
the solution for general relativistic  spherical accretion with and
without back reaction, and showed that relativistic effects enhance mass
accretion when back reaction  is neglected. The exact values of dynamical 
and thermodynamic accretion variables on the sonic surface,
and at extreme close vicinity of the black hole event horizons, have recently 
been calculated using complete general relativistic (Das 2002) as well 
as pseudo general relativistic (Das \& Sarkar 2001) treatments.\\
\noindent
For flow of matter with non-zero angular momentum density, spherical 
symmetry is broken and accretion phenomena is studied employing 
axisymmetric configuration. 
Accreting matter is thrown into circular orbits around the central accretor,
leading to the formation of the {\it accretion discs} around the galactic and
extra-galactic black holes.
For certain values of the intrinsic angular 
momentum density of accreting material, the number of sonic point, unlike spherical 
accretion, may exceed one, and accretion is called `multi-transonic'. Study of
such multi-transonicity was initiated by Abramowicz \& Zurek (1981). 
Subsequently, multi-transonic accretion disc in general relativity has been 
studied in a number of works (Fukue 1987, Chakrabarti 1990, 
Kafatos \& Yang 1994,
Yang \& Kafatos 1995, 
Pariev 1996, Peitz \& Appl 1997,    
Lu, Yu, Yuan \& Young 1997, Das 2004, Barai, Das \& Wiita 2004, hereafter 
BDW). All the above works, except BDW, usually deal with low angular 
momentum sub-Keplerian prograde flow. BDW studied the retrograde
flows as well and showed that a higher angular momentum (as high as
Keplerian) retrograde flow can also produce multi-transonicity. 
Sub-Keplerian 
weakly rotating flows
are exhibited in
various physical situations, such as detached binary systems
fed by accretion from OB stellar winds (Illarionov \&
Sunyaev 1975; Liang \& Nolan 1984), semi-detached low-mass
non-magnetic binaries (Bisikalo et al.\ 1998), and super-massive BHs fed
by accretion from slowly rotating central stellar clusters
(Illarionov 1988; Ho 1999 and references therein). Even for a standard Keplerian
accretion disk, turbulence may produce such low angular momentum flow
(e.g., Igumenshchev
\& Abramowicz 1999, and references therein).
\section{Motivation to study the analogue behaviour of transonic black hole accretion}
\noindent
Since the publication of the seminal paper by Bondi in 1952,
the transonic
behaviour of accreting fluid onto compact astrophysical objects has
been extensively studied in the astrophysics community, 
and the
pioneering work by Unruh in 1981 
initiated a substantial number of works
in the theory of {\AHR}  with diverse fields of application stated in \S 4.
It is surprising that no attempt was made to bridge
these two categories of research, astrophysical black hole accretion
and the theory of {\AHR},  by providing a self-consistent study of {\AHR}  for real
astrophysical fluid flows, i.e., by establishing the fact that
accreting black holes can be considered as a natural
example of analogue system. Since both the theory of transonic
astrophysical accretion and the theory of
{\AHR}  stem from almost
exactly the same physics, the propagation of a transonic fluid with
acoustic disturbances embedded into it, it is important to study
{\AHR} for transonic accretion onto astrophysical black
holes and to compute $T_{{AH}}$ for such accretion. \\
In the following sections, we will describe the details of
the transonic accretion and will show how the accreting black 
hole system can be considered as a classical analogue system.
Hereafter, we define gravitational radius
$r_g=\frac{2G{M_{BH}}}{c^2}$, where  $M_{BH}$  is the mass of the black
hole, $G$ is the universal gravitational constant and $c$ is the
velocity of light.
The radial distances and velocities are scaled in units of $r_g$ and $c$
respectively and all other derived quantities are scaled accordingly;
$G=c=M_{BH}=1$ is used.
\section{Newtonian and semi-Newtonian spherical accretion}
\noindent
In this section we will follow the methodology developed in 
Das \& Sarkar 2001 to describe the simplest possible example
of transonic black hole accretion.
If the gravitational potential is defined as $\Phi$, then the 
equation of motion for spherically accreting matter onto the 
black hole can be written as:
$$
\frac{{\partial{u}}}{{\partial{t}}}+u\frac{{\partial{u}}}{{\partial{r}}}+\frac{1}{\rho}
\frac{{\partial}p}{{\partial}r}+\frac{{\partial}\Phi}{\partial{r}}=0
\eqno{(9)}
$$
where symbols have their
usual meaning.
The first term of eqn. (9) is the Eulerian time derivative of the
dynamical velocity at a given $r$, the second term 
is the `advective' term, the third term 
is the
momentum deposition due to pressure gradient and the
final term is due to the gravitational acceleration.
Note that $\Phi$ may be purely Newtonian, or may represent 
any of the following semi-Newtonian pseudo-Schwarzschild 
black hole potentials:
$$
\Phi_{1}=-\frac{1}{2(r-1)},~ \Phi_{2}=-\frac{1}{2r}\left[1-\frac{3}{2r}+
12{\left(\frac{1}{2r}\right)}^2\right]
$$
$$
\Phi_{3}=-1+{\left(1-\frac{1}{r}\right)}^{\frac{1}{2}},~
\Phi_{4}=\frac{1}{2}ln{\left(1-\frac{1}{r}\right)}
\eqno{(10)}
$$
$\Phi_1$ is proposed by Paczy\'nski and Wiita (1980),
$\Phi_2$ by Nowak and Wagoner (1991), $\Phi_3$ and
$\Phi_4$ by Artemova, Bj\"{o}rnsson \&
Novikov (1996).
Introduction of such potentials allows one to investigate the
complicated physical processes taking place in accretion in a
semi-Newtonian framework by avoiding pure general relativistic calculations.
Most of the features of space-time around a compact object are retained and
some crucial properties of the analogous relativistic
accretion solutions could be reproduced with high accuracy.
Hence, those potentials might be designated as `pseudo-
Schwarzschild' potentials, see Das \& Sarkar 2001 and Das 2002 for details.\\
\noindent
The continuity
equation can be written as 
$$
\frac{{\partial}{\rho}}{{\partial}t}+\frac{1}{r^2}\frac{{\partial}}{{\partial}r}\left({\rho}ur^2\right)=0
\eqno{(11)}
$$
For a polytropic equation of state, the steady state
solutions 
of eqn. (9)  and eqn. (11) provides:\\
\noindent
1) Conservation of specific energy ${\cal E}$ of the flow:
$$
{\cal E}=\frac{u^2}{2}+\frac{a_s^2}{{\gamma}-1}+\Phi
\eqno{(12)}
$$
where $\gamma$(=$c_p/c_v$) is the adiabatic index of the accreting material.
\\
\noindent
2) Conservation of Baryon number (or accretion rate ${\dot M}$):
$$
{\dot M}=4{\pi}{\rho}ur^2
\eqno{(13)}
$$
Defining the entropy accretion rate $\Xi=
{\dot M}{\gamma}^{\frac{1}{\gamma-1}}K^{\frac{1}{\gamma-1}}$, we have:
$$
\Xi=4{\pi}a_s^{\frac{1}{\gamma-1}}ur^2
\eqno{(14)}
$$
The space
gradient of dynamical velocity is obtained as:
$$
{\left(\frac{du}{dr}\right)}=\frac{\frac{2a_s^2}{r}-
{\Phi}^{'}}{u-\frac{a_s^2}{u}}
\eqno{(15}
$$
where $\Phi^{'}$ is equal to $d{\Phi}/dr$.
Since the flow is assumed to be smooth everywhere, if
the denominator of eqn. (15)  vanishes at any radial distance
$r$, the numerator must also vanish there to maintain the
continuity of the flow.
One thus calculates
the sonic point quantities 
as
$$
{u}{\vh}=a_s{\vh}=\sqrt{\frac{r_h}{2}{{\Phi}^{'}}{{\Bigg{\vert}}}_{{\rh}}}
\eqno{(16)}
$$
where $r_h$ is the sonic point, a family of which forms the spherical 
acoustic horizon.
The location of the acoustic horizon, i.e.,
$r_h$, can be obtained
by algebraically solving the following equation:
$$
{\cal E}-\frac{1}{2}\left(\frac{\gamma+1}{\gamma-1}\right)r_h
{{\Phi}^{'}}{{\Bigg{\vert}}_{\rh}}
-{\Phi}{\Bigg{\vert}}_{\rh}=0
\eqno{(17)}
$$
$\left(\frac{du}{dr}\right)_{\rh}$ 
at its
corresponding sonic point can be obtained by
solving the following quadratic equation:
$$
\left(1+\gamma\right){{\left(\frac{du}{dr}\right)}^2}_{\rh}+
2.829\left(\gamma-1\right)\sqrt\frac{{{\Phi}^{'}}
{{\Bigg{\vert}}_{\rh}}}{r_h}{\left(\frac{du}{dr}\right)}_{\rh}
+\left(2{\gamma}-1\right)
\frac{
{{\Phi}^{'}{\Bigg{\vert}}_{\rh}}}
{{r_h}}
+{{\Phi}^{''}}{{\Bigg{\vert}}_{\rh}}=0
\eqno{(18)}
$$
\noindent
From the above discussion it is evident that for completely Newtonian flow
($\Phi=-\frac{1}{r}$), as well as for semi-Newtonian flow, one obtains the 
{\it exact} location of the transonic point, which is the location of the 
acoustic horizon because of the stationarity approximation. One can
also calculate the {\it exact} value of $\left[u,a_s,du/dr,da_s/dr\right]$
at the acoustic horizon. From eq. (8), it is obvious that all the required 
quantities to calculate the value of analogue Hawking temperature may be 
obtained from first principle from the calculations presented above, if the
gravitational potential is purely Newtonian in nature. Hence, Newtonian 
spherical accretion onto astrophysical black holes {\it is} the example
of analogue system where the {\it exact} value of $T_{AH}$ can be calculated
using {\it only two measurable parameters}, namely, ${\cal E}$ and $\gamma$.
Also to be noted that the above formulation was carried out using 
{\it position dependent} sound speed. Unruh's original calculation of $T_{{AH}}$ 
(see eq. (4)) assumes the
positional constancy of sound speed, which may not always
be a real physical situation.
This was modified to include the position dependent sound speed by
Visser (1998) for flat space and by Bili$\acute{c}$ (1999)
for curved space. However, none of these
works provides the exact calculation of $r_h$ and {\AH} related quantities
from  first principles, and hence the numerical value of $T_{{AH}}$ in {\it any}
existing literature (even for flat space, let alone curved space) was
obtained using approximate, order of magnitude, calculations only.\\
\noindent
The most important point to mention in this context is that,
the accreting astrophysical black holes are the {\it only}
real physical candidates for which both the horizons, actual (electromagnetic)
and analogue (sonic), may exist together {\it simultaneously}. Hence
our application of {\AHR} to the theory of transonic
astrophysical accretion may be useful to compare
some properties of these two kind of horizons,
by theoretically studying and comparing the behaviour of the {\it same} flow
close to these horizons. 
Spherically accreting black hole system is, therefore, not only a
concrete example of analogue system naturally found in the Universe, it is
also {\it most important} among all the existing analogue models, because 
study of black hole accretion as black hole analogue may bridge the gap
between two apparently different school of thoughts, the theory of transonic
astrophysical accretion, and the theory of analogue Hawking radiation, 
sharing intrinsically the same physical origin.\\
\noindent
If $\Phi$ represents any of the pseudo-Schwarzschild potentials defined by
eq. (10), the analogue Hawking temperature can still be determined from
the knowledge of eq. (16 - 18). One only needs to modify 
eq. (8) by incorporating such potentials instead of the purely Newtonian one, while
calculating the acoustic metric and other related quantities. It is not difficult to 
show that (Das \& Dasgupta 2004, in preparation) for purely Newtonian as well as for 
semi-Newtonian black hole potentials, $T_{AH}$ can never exceed $T_{H}$, which is
not true for general relativistic accretion, as we will see in \S 10.
\section{General relativistic black hole accretion}
In this section, we will describe fully general relativistic astrophysical accretion,
both spherically symmetric and axisymmetric, onto astrophysical black holes. 
The mass of the
accreting fluid is assumed to be much less compared to $M_{BH}$ (which is usually
reality for astrophysical black hole accretion), so that the gravity field is
controlled essentially by $M_{BH}$ only. Calculations presented in the following sections
are mostly adopted from Das 2004, Das 2004a and BDW.
\subsection{Mono-transonic spherical accretion}
\noindent
Accretion is $\theta$ and $\phi$ symmetric and possesses only radial inflow velocity. 
Stationary solutions are considered. Accretion is governed by the radial part
of the general relativistic 
time independent Euler and continuity equations in Schwarzschild
metric. The conserved specific flow energy ${\cal E}$ (the relativistic 
analogue of Bernoulli's constant) along each stream line reads ${\cal E}=hu_t$ 
(Anderson 1989) where
$h$ and $u_\mu$ are the specific enthalpy and the four velocity, which can be 
re-cast in terms of the radial three velocity $u$ and the polytropic sound speed 
$a_s$ to obtain:
$$
{\cal E}=\left[\frac{\gamma-1}{\gamma-\left(1+a^2_s\right)}\right]
\sqrt{\frac{1-\frac{1}{r}}{1-u^2}}.
\eqno{(19)}
$$
We concentrate on positive Bernoulli constant solutions.
The mass accretion rate ${\dot M}$ may be obtained by integrating the continuity
equation:
$$
{\dot M}=4{\pi}{\rho}ur^2\sqrt{\frac{r-1}{r\left(1-u^2\right)}},
\eqno{(20)}
$$
where $\rho$ is the proper mass density.  A polytropic equation of state, $p=K\rho^\gamma$,  is 
assumed (this is common in astrophysics
to describe relativistic accretion) to define $\Xi$ as:
$$
\Xi=K^{\frac{1}{\gamma-1}}{\dot M}=4{\pi}{\rho}ur^2\sqrt{\frac{r-1}{r\left(1-u^2\right)}}
\left[\frac{a^2_s\left(\gamma-1\right)}{\gamma-\left(1+a^2_s\right)}\right].
\eqno{(21)}
$$
Simultaneous solution of eq. (19 - 21) provides the dynamical three velocity gradient 
at any radial distance $r$:
$$
\frac{du}{dr}=\frac{u\left(1-u^2\right)\left[a^2_s\left(4r-3\right)-1\right]}
{2r\left(r-1\right)\left(u^2-a^2_s\right)}
\eqno{(22)}
$$
The sonic point conditions are obtained as:
$$
u{\vh}=a_s{\vh}=\sqrt{\frac{1}{4r_h-3}}.
\eqno{(23)}
$$
Substitution of $u{\vh}$ and $a_s{\vh}$ into eq. (19) for $r=r_h$ provides:
$$
r_h^3+r_h^2\Gamma_1+r_h\Gamma_2+\Gamma_3=0,
\eqno{(24)}
$$
where
$$
\Gamma_1=\left[\frac{2{\cal E}^2\left(2-3\gamma\right)+9\left(\gamma-1\right)}
         {4\left(\gamma-1\right)\left({\cal E}^2-1\right)}\right],~
$$
$$
\Gamma_2=\left[\frac{{\cal E}^2\left(3\gamma-2\right)^2-
          27\left(\gamma-1\right)^2}
          {32\left({\cal E}^2-1\right)\left(\gamma-1\right)^2}\right],~
\Gamma_3=\frac{27}{64\left({\cal E}^2-1\right)}.
\eqno{(25)}
$$
Solution of eq. (24) provides the location of the {\AH} in terms of only two accretion parameters
$\{{\cal E},\gamma\}$, which is the two parameter input set to study the flow.
Eq. (24) can be solved completely analytically 
by employing the Cardano-Tartaglia-del Ferro technique. One defines:
$$
\Sigma_1=\frac{3\Gamma_2-\Gamma_1^2}{9},~
\Sigma_2=\frac{9\Gamma_1\Gamma_2-27\Gamma_3-2\Gamma_1^3}{54},~
\Psi=\Sigma_1^3+\Sigma_2^2,~
$$
$$
 \Theta={\rm cos}^{-1}\left(\frac{\Sigma_2}{\sqrt{-\Sigma_1^3}}\right)
\Omega_1=\sqrt[3]{\Sigma_2+\sqrt{\Sigma_2^2+\Sigma_1^3}},~
$$
$$
\Omega_2=\sqrt[3]{\Sigma_2-\sqrt{\Sigma_2^2+\Sigma_1^3}},~
\Omega_{\pm}=\left(\Omega_1\pm\Omega_2\right)
\eqno{(26)}
$$
so that the three roots for $r_h$ come out to be:
$$
^1\!r_h=-\frac{\Gamma_1}{3}+\Omega_+,~
^2\!r_h=-\frac{\Gamma_1}{3}-\frac{1}{2}\left(\Omega_+-i\sqrt{3}\Omega_-\right),~
$$
$$
^3\!r_h=-\frac{\Gamma_1}{3}-\frac{1}{2}\left(\Omega_--i\sqrt{3}\Omega_-\right).
~\eqno{(27)}
$$
However, note that not all $^i\!r_h\{i=1,2,3\}$ would be real for all \egam. It is
easy to show that if $\Psi>0$, only one root is real; if $\Psi=0$, all roots are
real and at least two of them are identical; and if $\Psi<0$, all roots are real 
and distinct. Selection of the real physical ($r_h$ has to be greater than unity) roots
requires
the following discussion. \\ 
\noindent
Although one can analytically calculate $r_h$ and other variables at $r_h$, there is no 
way that one can analytically calculate the flow variables at any arbitrary $r$. One needs
to integrate eq. (24) numerically to obtain the transonic profile of the flow for 
all range of $r$, starting from infinity and ending on to the actual event horizon. To do
so, one must set the appropriate limits on \egam  to model the realistic situations
encountered in astrophysics. As ${\cal E}$ is scaled in terms of
the rest mass energy and includes the rest mass energy, 
${\cal E}<1$ corresponds to the negative energy accretion state where
radiative extraction of rest mass energy from the fluid is required. For such extraction 
to be made possible, the accreting fluid has to
possess viscosity or other dissipative mechanisms, which may violate the Lorenzian invariance.
On the other hand, although almost any ${\cal E}>1$ is mathematically allowed, large 
values of ${\cal E}$ represents flows starting from infinity 
with extremely high thermal energy, and ${\cal E}>2$ accretion represents enormously 
hot flow configurations at very large distance from the black hole, 
which are not properly conceivable in realistic astrophysical situations.
Hence one sets $1{\lsim}{\cal E}{\lsim}2$. Now, $\gamma=1$ corresponds to isothermal accretion
where accreting fluid remains optically thin. This is the physical lower limit for 
$\gamma$; $\gamma<1$ is not realistic in accretion
astrophysics. On the other hand,
$\gamma>2$ is possible only for superdense matter 
with substantially large magnetic
field (which requires the accreting material to be governed by general relativistic 
magneto-hydrodynamic 
equations, dealing with which
is beyond the scope of this article) and direction dependent anisotropic pressure. One thus 
sets $1{\lsim}\gamma{\lsim}2$ as well, so \egam has the boundaries
$1{\lsim}\{{\cal E},\gamma\}{\lsim}2$. However, one should note that the most preferred 
values of $\gamma$ for realistic black hole accretion ranges from $4/3$ 
to $5/3$ (Frank et. al. 1992).\\
\noindent
Coming back to the solution for $r_h$, one finds that for the preferred range of \egam,
one {\it always} obtains $\Psi<0$. Hence the roots are always real and three real
unequal roots can be computed as:
$$
^1\!{{r}}_h=2\sqrt{-\Sigma_1}{\rm cos}\left(\frac{\Theta}{3}\right)
                  -\frac{\Gamma_1}{3},
^2\!{{r}}_h=2\sqrt{-\Sigma_1}{\rm cos}\left(\frac{\Theta+2\pi}{3}\right)
                  -\frac{\Gamma_1}{3},~
$$
$$
^3\!{{r}}_h=2\sqrt{-\Sigma_1}{\rm cos}\left(\frac{\Theta+4\pi}{3}\right)
                  -\frac{\Gamma_1}{3}
\eqno{(28)}
$$
One finds that for {\it all} $1{\lsim}${\egam}${\lsim}2$, $^2\!{{r}}_h$ becomes negative. 
It is observed 
that $\{^1\!{{r}}_h,^3\!{{r}}_h\}>1$ for most values of the astrophysically 
tuned \egam.
However, it is also found that $^3\!{{r}}_h$ does not allow steady physical flows to pass
through it; either $u$, or $a_s$, or both, becomes superluminal before the flow reaches
the actual event horizon, or the Mach number profile shows intrinsic fluctuations for 
$r<r_h$. This information is obtained by numerically integrating the 
complete flow profile passing through $^3\!{{r}}_h$. Hence it turns out that one needs to
concentrate {\it only} on  $^1\!{{r}}_h$ for realistic astrophysical black hole accretion. 
Both large ${\cal E}$ and large $\gamma$ enhance the thermal energy of the flow 
so that the 
accreting fluid acquires the large radial velocity to overcome $a_s$ only in the 
close vicinity of the black hole . Hence $r_h$ anti-correlates with \egam. 
To obtain
$(dv/dr)$ and $(da_s/dr)$ on the {\AH}, L' Hospital's rule is applied to 
eq. (22) to have:
$$
\left(\frac{du}{dr}\right)_{r=r_h}=\Phi_{12}-\Phi_{123},~
$$
$$
\left(\frac{da_s}{dr}\right)_{r=r_h}=\Phi_4\left(\frac{1}{\sqrt{4r_h-3}}+\frac{\Phi_{12}}{2}
                            -\frac{\Phi_{123}}{2}\right)
\eqno{(29)}
$$
where
$$
\Phi_{12}=-\Phi_2/2\Phi_1,~
\Phi_{123}=\sqrt{\Phi_2^2-4\Phi_1\Phi_3}/2\Phi_1,~
$$
$$
\Phi_1=\frac{6r_h\left(r_h-1\right)}{\sqrt{4r_h-3}},~
\Phi_2=\frac{2}{4r_h-3}\left[4r_h\left(\gamma-1\right)-
        \left(3\gamma-2\right)\right]
$$
$$
\Phi_3=\frac{8\left(r_h-1\right)}{\left(4r_h-3\right)^{\frac{5}{2}}}
       [r_h^2\left(\gamma-1\right)^2-r_h\left(10\gamma^2-19\gamma+9\right)
$$
$$
       +\left(6\gamma^2-11\gamma+3\right)],~
\Phi_4=\frac{2\left(2r_h-1\right)-\gamma\left(4r_h-3\right)}
       {4\left(r_h-1\right)}
\eqno{(30)}
$$
\subsection{Multi-transonic disc accretion}
In this section, we follow the treatment provided by Das 2004 and BDW because 
only in those works a consistent methodology has been developed which can handle
the transonic solutions upto the {\it extreme close} vicinity 
of the black hole event horizon (down to a distance upto around $1.001 r_g$).
The Boyer--Lindquist
coordinates with signature $-+++$ is used, and an
azimuthally Lorentz boosted orthonormal tetrad basis co-rotating
with the accreting fluid. $\lambda$ is defined to be the specific
angular momentum of the flow and any gravo-magneto-viscous
non-alignment between $\lambda$, and the Kerr parameter $a$, is neglected.
Stationary
axisymmetric solution of the following equations is considered:
$$
{\bf \nabla}^\mu{\Im}^{{\mu}{\nu}}
=0,~
~
\left({\rho}{v^\mu}\right)_{;\mu}=0,
\eqno{(31)}
$$
\noindent
where ${\Im}^{{\mu}{\nu}}, v_\mu$, and $\rho$ are the energy momentum tensor, the four
velocity and the rest mass density of the
accreting fluid, respectively. The semicolon  denotes the
covariant derivative. 
The `stationarity' condition implies the vanishing of the
temporal derivative of any scalar field in the disc or the
vanishing of the Lie Derivative of any vector or tensor field
along the Killing vector $\frac{\partial}{{\partial}t}$.
This implies that 
$\frac{{\partial}{\Pi}}{{\partial}r}=0$ if ${\Pi}$ is a scalar
field, and ${\cal L}_{{\partial}/{\partial}t}{\bf {\Pi}}=0$ if
${\bf {\Pi}}$ represents a vector or a tensor field.
The `axisymmetry' property is endowed with the space like
Killing field
$\left(\frac{\partial}{{\partial}\phi}\right)^\mu$.\\
\noindent
The radial momentum balance condition
may be obtained from
$\left(v_{\mu}v_{\nu}+g_{{\mu}{\nu}}\right){\Im}^{{\mu}{\nu}}_{;{\nu}} = 0$.
Exact solutions of these conservation equations
require knowledge of the accretion
geometry and the introduction of a suitable equation of
state.
Polytropic accretion is considered for which
$p=K{\rho}^{\gamma}$, where $K$ and $\gamma$
are the monotonic and continuous
functions of the specific entropy density, and  the constant
adiabatic index of the flow, respectively.
The specific proper flow enthalpy is taken to be
$h=\left(\gamma-1\right)\left\{{\gamma}-\left(1+a_s^2\right)\right\}^{-1}$,
where $a_s$ is the polytropic sound speed defined as
$a_s=\left(\partial{p}/\partial{\epsilon}\right)_{\cal S}^{1/2}
={\bf \Psi_1}\left(T(r),\gamma\right)={\bf \Psi_2}\left(p,{\rho},\gamma\right)$;
here $T(r)$ is the local flow temperature,
$\epsilon$ is the mass-energy density,
and $\left\{{\bf \Psi_1},{\bf \Psi_2}\right\}$ are known
functions. The subscript ${\cal S}$ indicates that the derivative
is taken at constant specific entropy. Hence,
$a_s^2(r)= \gamma{\kappa}T(r)/({\mu}m_H)={\Theta}^2T(r)$,
where $\Theta = [\gamma{\kappa}/(\mu{m_H})]^{1/2}$,
$\mu$ is the mean molecular weight,
$m_H$ is the mass of the hydrogen atom,
and $\kappa$ is Boltzmann's constant.\\
\noindent
The temporal component of the first part of eq. (31) leads to
the conservation of specific flow energy ${\cal E}$.
The Kerr metric in the equatorial
plane of the black hole may be written 
as (e.g., Novikov \& Thorne 1973)
$$
ds^2=g_{{\mu}{\nu}}dx^{\mu}dx^{\nu}=-\frac{r^2{\Delta}}{A}dt^2
+\frac{A}{r^2}\left(d\phi-\omega{dt}\right)^2
+\frac{r^2}{\Delta}dr^2+dz^2 ,
\eqno{(32)}
$$
\noindent
where
$\Delta=r^2-2r+a^2, ~A=r^4+r^2a^2+2ra^2$, 
and $\omega=2ar/A$.
The angular velocity $\Omega$ is
$$
{\Omega}=-\frac{\left(g_{t\phi}+\lambda g_{tt}\right)}
{\left(g_{{\phi}\phi}+\lambda g_{t\phi}\right)}
=\frac{\frac{4a}{r}-\frac{\lambda \left(4a^2-r^2\Delta \right)}{A}} {\frac{A}{r^2}-
\frac{4{\lambda}a}{r}}.
\eqno{(33)}
$$
The normalization relation $v_\mu{v^\mu}=-1$, along with the above 
value of $\Omega$,  
provides:
$$
v_t=
\left[\frac{Ar^2\Delta}
{\left(1-u^2\right)\left\{A^2-8\lambda arA
+\lambda^2r^2\left(4a^2-r^2\Delta\right)\right\}}\right]^\frac{1}{2},
\eqno{(34)}
$$
\noindent
where $u$ is the radial three velocity in the co-rotating fluid frame.
Hence 
the conserved specific energy (which includes the rest mass energy) 
can be re-written as:
$$
{\cal E}=\frac{\left(\gamma-1\right)}
{\left[\gamma-\left(1+\Theta^2T\right)\right]}
\left[\frac{Ar^2\Delta}
{\left(1-u^2\right)\left\{A^2-8\lambda arA
+\lambda^2r^2\left(4a^2-r^2\Delta\right)\right\}}\right]^\frac{1}{2}
\eqno{(35)}
$$
For a suitably chosen disc geometry (see BDW for details about the disk structure
in Kerr metric,
also see Das 2004, for the details of 
relativistic accretion in a different disc geometry in Schwarzschild metric)
The mass and entropy accretion rate comes out to be:
$$
{\dot M}=4{\pi}{\rho}{\Theta}T^{\frac{1}{2}}Mr^2
\left[\frac{2\Delta\left(\gamma-1\right)\Theta^2T}
           {\gamma \left(1-\Theta^2TM^2\right) \left\{\gamma-\left(1+\Theta^2T\right)\right\}
            \psi}\right]^{\frac{1}{2}};
\eqno{(36)}
$$
$$
{\dot {\bf \Xi}}=
4\sqrt{2}\pi r^2 u
\sqrt{ \frac{\Delta} {\left(1-u^2\right)\psi} }
\left[\frac{a_{s}^2\left(\gamma-1\right)}{\gamma\left\{\gamma-\left(1+a_{s}^2\right)\right\}}\right]
^{{\frac{1}{2}}\left(\frac{\gamma+1}{\gamma-1}\right)}.
\eqno{(37)}
$$
The velocity gradient can be calculated as:
$$
\frac{du}{dr}=
\frac
{ \frac{2a_{s}^2}{\left(\gamma+1\right)}
  \left[ \frac{r-1}{\Delta} + \frac{2}{r} -
         \frac{v_{t}\sigma \chi}{4\psi}
  \right] -
  \frac{\chi}{2} }
{ \frac{u}{\left(1-u^2\right)} -
  \frac{2a_{s}^2}{ \left(\gamma+1\right) \left(1-u^2\right) u }
   \left[ 1-\frac{u^2v_{t}\sigma}{2\psi} \right] },
\eqno{(38)}
$$
\noindent where
$$
\sigma = 2\lambda^2v_{t}-a^2,
~
\chi =
\frac{1}{\Delta} \frac{d\Delta}{dr} +
\frac{\lambda}{\left(1-\Omega \lambda\right)} \frac{d\Omega}{dr} -
\frac{\left( \frac{dg_{\phi \phi}}{dr} + \lambda \frac{dg_{t\phi}}{dr} \right)}
     {\left( g_{\phi \phi} + \lambda g_{t\phi} \right)}.
\eqno{(39)}
$$
\noindent
The sonic point conditions can be obtained as:
$$
a_{s}{\vert}_{\rh}={\left[\frac{u^2\left(\gamma+1\right)\psi}
                                  {2\psi-u^2v_t\sigma}
                       \right]^{\frac{1}{2}}_{\rh}  },
~~u{\vh}= {\left[\frac{\chi\Delta r} {2r\left(r-1\right)+ 4\Delta} \right]
^{\frac{1}{2}}_{\rh}  },
\eqno{(40)}
$$
\noindent
For any value of \eker, 
substitution of the values of $u{\vh}$ and $a_{s}{\vert}_{\rh}$  in terms of $r_h$ 
in the expression 
for ${\cal E}$ (see eq. (35)),
provides
a polynomial in $r_h$, the solution of which determines
the location of the sonic point(s) $r_h$.
After lengthy algebraic manipulations,
the following quadratic equation 
is formed,
which can be solved 
to obtain $(du/dr)_{\rh}$:
$$
\alpha \left(\frac{du}{dr}\right)_{\rh}^2 + \beta \left(\frac{du}{dr}\right)_{\rh} + \zeta = 0,
\eqno{(41)}
$$
\noindent
where the coefficients are 

$$
\alpha=\frac{\left(1+u^2\right)}{\left(1-u^2\right)^2} - \frac{2\delta_1\delta_5}{\gamma+1},
~~~\beta=\frac{2\delta_1\delta_6}{\gamma+1} + \tau_6,
~~~\zeta=-\tau_5;
$$
$$
\delta_1=\frac{a_s^2\left(1-\delta_2\right)}{u\left(1-u^2\right)},
~\delta_2 = \frac{u^2 v_t \sigma}{2\psi},
~\delta_3 = \frac{1}{v_t} + \frac{2\lambda^2}{\sigma} - \frac{\sigma}{\psi} ,
$$
$$
\delta_4 = \delta_2\left[\frac{2}{u}+\frac{u v_t \delta_3}{1-u^2}\right],
~
\delta_5 = \frac{3u^2-1}{u\left(1-u^2\right)} - \frac{\delta_4}{1-\delta_2} -
           \frac{u\left(\gamma-1-a_s^2\right)}{a_s^2\left(1-u^2\right)},
$$
$$
\delta_6 = \frac{\left(\gamma-1-a_s^2\right)\chi}{2a_s^2} +
           \frac{\delta_2\delta_3 \chi v_t}{2\left(1-\delta_2\right)},
$$
$$
\tau_1=\frac{r-1}{\Delta} + \frac{2}{r} - \frac{\sigma v_t\chi} {4\psi},
~
\tau_2=\frac{\left(4\lambda^2v_t-a^2\right)\psi - v_t\sigma^2} {\sigma \psi},
$$
$$
\tau_3=\frac{\sigma \tau_2 \chi} {4\psi},
~
\tau_4 = \frac{1}{\Delta} 
       - \frac{2\left(r-1\right)^2}{\Delta^2}
       -\frac{2}{r^2} - \frac{v_t\sigma}{4\psi}\frac{d\chi}{dr},
$$
$$
\tau_5=\frac{2}{\gamma+1}\left[a_s^2\tau_4 -
     \left\{\left(\gamma-1-a_s^2\right)\tau_1+v_ta_s^2\tau_3\right\}\frac{\chi}{2}\right]
   - \frac{1}{2}\frac{d\chi}{dr},
$$
$$
\tau_6=\frac{2 v_t u}{\left(\gamma+1\right)\left(1-u^2\right)}
       \left[\frac{\tau_1}{v_t}\left(\gamma-1-a_s^2\right) + a_s^2\tau_3\right].
\eqno{(42)}
$$
Note that all the above quantities are evaluated at $r_h$.\\
\noindent
Putting the Kerr parameter equals to zero in the corresponding equations 
above, one can obtain the value of sonic point and $\left[u,a_s,du/dr,da_s/dr\right]$
at the sonic point, for accretion onto Schwarzschild black holes.\\
\noindent 
Adopting a different disc geometry,
the corresponding quantities in Schwarzschild metric can also directly be 
computed as the following (Das 2004):\\
\noindent
1) The sonic point is obtained by solving the following equation:
$$
{\cal  E}^2\left[r_h^3+\lambda^2\left(1-r_h\right)\right]
-\frac{r_h-1}{1-{\bf {{\Psi}\left(r_h,\lambda\right)}}}
\left[\frac{r_h\left(\gamma-1\right)}{\gamma-
{\bf {{\eta}\left(r_h,\lambda\right)}}}\right]^2=0
\eqno{(43)}
$$
where
$$
{\bf {{\eta}\left(r_h,\lambda\right)}}
=\left[1+\frac{\gamma+1}{2}{\bf
{{\Psi}\left(r_h,\lambda\right)}}\right]
$$
and
$$
{\bf {{\Psi}\left(r_h,\lambda\right)}}=
\left[
\frac{
\frac{1}{2r_h}\left[\frac{2r_h^3-\lambda^2}{r_h^3+\lambda^2\left(1-r\right)}
\right]
-\frac{2r_h-1}{2r_h\left(r_h-1\right)}
}
{\frac{1-2r_h}{2r\left(r_h-1\right)}
+\frac{\lambda^2-3r_h^2}{2\left[r_h^3+\lambda^2\left(1-r_h\right)\right]}
}
\right]
$$
\noindent
2) The sonic velocities can be found as:\\
$$
u{\vh}=\sqrt{\frac{2}{\gamma+1}}a_s{\vh}=
\sqrt{
\left[
\frac{
\frac{1}{2r}\left[\frac{2r^3-\lambda^2}{r^3+\lambda^2\left(1-\gamma\right)}
\right]
-\frac{2r-1}{2r\left(r-1\right)}
}
{\frac{1-2r}{2r\left(r-1\right)}
+\frac{\lambda^2-3r^2}{2\left[r^3+\lambda^2\left(1-r\right)\right]}
}
\right]_{\rh}}
\eqno{(44)}
$$
3) The velocity gradient can be obtained by solving the following equation:\\
$$
\frac{2\left(2\gamma-3a_h^2\right)}
{\left(\gamma+1\right)\left(u_h^2-1\right)^2}
\left(\frac{du}{dr}\right)_s^2+
4{\bf {\xi\left(r_h,\lambda\right)}}
\left[\frac{\gamma-\left(1+a_h^2\right)}{u_h^2-1}\right]
\left(\frac{du}{dr}\right)_h
$$
$$
+ \frac{2}{\gamma+1}a_h^2{\bf {\xi\left(r_h,\lambda\right)}}
{\Bigg [}
2{\bf {\xi\left(r_h,\lambda\right)}}\left[
\frac{\gamma-\left(1+a_h^2\right)}{\gamma+1}\right]-
\frac{2r_h-1}{r_h\left(r_h-1\right)}
-\frac{3r_h^2-\lambda^2}{r_h^3+\lambda^2\left(1-r_h\right)}
$$
$$
+\frac{20r_h^3-12r_h^2-2\lambda^2\left(3r_h-2\right)}
{5r_h^4-4r_h^3-\lambda^2\left(3r_h^2-4r_h+1\right)}
{\Bigg  ]}=0
\eqno{(45)}
$$
where
$$
{\bf {\xi\left(r_h,\lambda\right)}}=
\left[
\frac{1-2r}{2r\left(r-1\right)}
+\frac{\lambda^2-3r^2}{2\left[r^3+\lambda^2\left(1-r\right)\right]}
\right]_{r=r_h}
$$
In the above two equations, $u_h$ and $a_h$ indicates the value of 
$u$ and $a_s$ on the acoustic horizon (on $r=r_h$), and the subscript
$h$ in general indicates that the quantities are measured on the 
acoustic horizon.
The relation between the dynamical velocity gradient and acoustic velocity gradient at 
any point (including at the sonic point) can be obtained as:\\
 $$
 \frac{da_s(r)}{dr}=
 \frac{a_s}{\gamma+1}\left\{\gamma-\left(1+a^2_s\right)\right\}
 \left[\frac{1}{u\left(u^2-1\right)}\frac{du}{dr}+
 {\bf f}\left(r,\lambda\right)\right]
 \eqno{(46)}
 $$
where:
$$
{\bf f}\left(r,\lambda\right)=
\frac{1-2r}{2r\left(r-1\right)}+
\frac{\lambda^2-3r^2}{2\left(r^3-\lambda^2r+\lambda^2\right)}
\eqno{(47)}
$$
Note that unlike spherical accretion, one finds three sonic points in disc accretion for 
some values of $\left[{\cal E},\lambda,\gamma\right]$ for accretion onto the
Schwarzschild as well as 
the Kerr black hole,
among which the largest and the smallest values correspond to the $X$ type outer, $r_o$, and
inner, $r_i$, sonic points, respectively. The $O$ type middle sonic point, $r_m$, which is
unphysical in the sense that no steady transonic solution passes through it, lies
between $r_i$ and $r_o$. 
If the accretion through $r_o$ can be perturbed in a
way so that it produces an amount of entropy exactly equal
to $\left[\Xi(r_i)-\Xi(r_o)\right]$,
supersonic flow through $r_o$ can join the subsonic flow through $r_i$ by developing
a standing shock. The exact location of such a shock, as well as the details of the post-shock
flow, may be obtained by formulating and solving the general relativistic Rankine-Hugoniot
condition for the flow geometry described above. 
Note that as a flow through $r_i$ does not connect
the black hole event horizon 
with infinity (such a flow folds back onto itself, see, e.g.,
 Fig. 1 of BDW), solutions through
$r_i$ do not have {\it independent} physical existence, and can only be accessed
if the supersonic flow through $r_o$ undergoes a shock, so that it generates extra
entropy, becomes subsonic, and produces the physical segment of the solution through $r_i$.
\section{Analogue Hawking temperature for general relativistic accretion}
\subsection{Relativistic sonic geometry}
\noindent
As mentioned in \S 8, it is not quite difficult
to calculate $T_{AH}$ for Newtonian, and also, perhaps, for semi-Newtonian 
black hole accretion. For general relativistic flow, however, non-trivial 
effort has to be made to accomplish such goals. This is because, treatments
in Unruh (1981) and Visser (1998), and in related works, are based on the
distinct fact that the background fluid is governed by Newtonian spacetime, whereas
for general relativistic accretion presented in \S 9.1 \& 9.2 in this paper, the
background fluid is characterized by purely Schwarzschild metric. Hence a
relativistic version of acoustic geometry is necessary to calculate $T_{AH}$ for
this purpose. The work by Bilic (1999) is the first (and perhaps, the only one
in the literature as far as our knowledge is concerned) which calculates the
generalized expression for acoustic analogue of surface gravity for propagation of
relativistic transonic fluid. The relativistic 
acoustic metric $G_{\mu{\nu}}$ comes out to be 
(Moncrief 1980, Bilic 1999):
$$
G_{\mu{\nu}}=\frac{N}{h{a_s}}\left[g_{\mu{\nu}}-\left(1-a_s^2\right)
u_{\mu}u_{\nu}\right]
\eqno{(48)}
$$
where $h$ is the relativistic enthalpy density and $N$ is the 
particle number density. 
The corresponding surface gravity may be calculated as:
$$
\kappa=\left|\frac{\sqrt{{\xi}^{\nu}{\xi}_{\nu}}}{1-a_s^2}
\frac{\partial}{\partial{n}}
\left(u-a_s\right)\right|_{\rm acoustic~ horizon}
\eqno{(49)}
$$
Where $\xi^{\mu}$ is the stationary Killing field and $\partial/{\partial{n}}$ is the 
normal derivative.
\subsection{Analogue temperature for spherical accretion}
\noindent
For spherically symmetric general relativistic flow onto Schwarzschild black holes
described in \S 9.1, one can evaluate the exact value of the Killing fields and 
Killing vectors to calculate the surface gravity for that geometry. The
analogue Hawking temperature for such geometry comes out to be (Das 2004a):
$$
T_{{AH}}=\frac{\hbar{c^3}}{4{\pi}{k_b}GM_{BH}}
    \left[\frac{r_h^{\frac{1}{2}}\left(r_h-0.75\right)}
    {\left(r_h-1\right)^{\frac{3}{2}}}\right]
    \left|\frac{d}{dr}\left(a_s-u\right)\right|_{r=r_h},
\eqno{(50)}
$$
where the values of $r_h, (du/dr)_h$ and $(da_s/dr)_h$ are obtained using the system of
units and scaling used in this article.\\
It is evident from the above formula that the {\it exact} value of $T_{AH}$ can
be {\it analytically} calculated from the results obtained in \S 9.1. While eq. (27)
provides the location of the acoustic horizon ($r_h$), the value of 
$\left|\frac{d}{dr}\left(a-u\right)\right|_{r=r_h}$ is obtained from eq. (29 - 30)
as a function of ${\cal E}$ and $\gamma$, both of which are real, physical,
measurable quantities. 
Note again, that, since $r_h$ and other quantities
appearing in eq. (50) are {\it analytically} calculated as a function of
\egam, eq. (50) provides an {\it exact analytical value} of the
general relativistic analogue Hawking temperature for {\it all possible solutions} of an
spherically accreting
astrophysical black hole system, something which has never been done in the literature
before.
If $\sqrt{4r_h-3}(1/2-1/\Phi_4)(\Phi_{12}-\Phi_{123})>1$,
one {\it always} obtains $(da_s/dr<du/dr)_h$ from eq. (29), which indicates the presence of the
acoustic white holes at $r_h$. This inequality holds good for
certain astrophysically relevant range of \egam; see following discussions.\\
\noindent
For a particular value of \egam, one can define the quantity $\tau$ to be the ratio of
$T_{{AH}}$ and $T_H$ as:
$$
\tau=T_{{AH}}/T_H.
\eqno{(51)}
$$
This ratio $\tau$ comes out to be independent of the mass of the black hole,
which enables us to compare the properties of two kind of horizons (actual and acoustic) for
an spherically
accreting black hole with any mass,
starting from the primordial holes
to the super massive black holes
at galactic centers. \\
\noindent
Note, however, that the analogue has been applied to
describe the classical perturbation of the fluid in terms of a
field satisfying the wave equation on a curved effective geometry. 
Main motivation of the methodology described in this 
section is not a rigorous demonstration of how
the phonon field generated in this system could be made quantized. To
accomplish that task, one can show that the effective action for the
acoustic perturbation is equivalent to the field theoretical action
in curved space, and the corresponding commutation and dispersion
relations (see, e.g., Unruh \& Sch$\ddot{\rm u}$tzhold 2003) may directly follow from there.\\
\begin{figure}
\vbox{
\vskip -4.5cm
\centerline{
\psfig{file=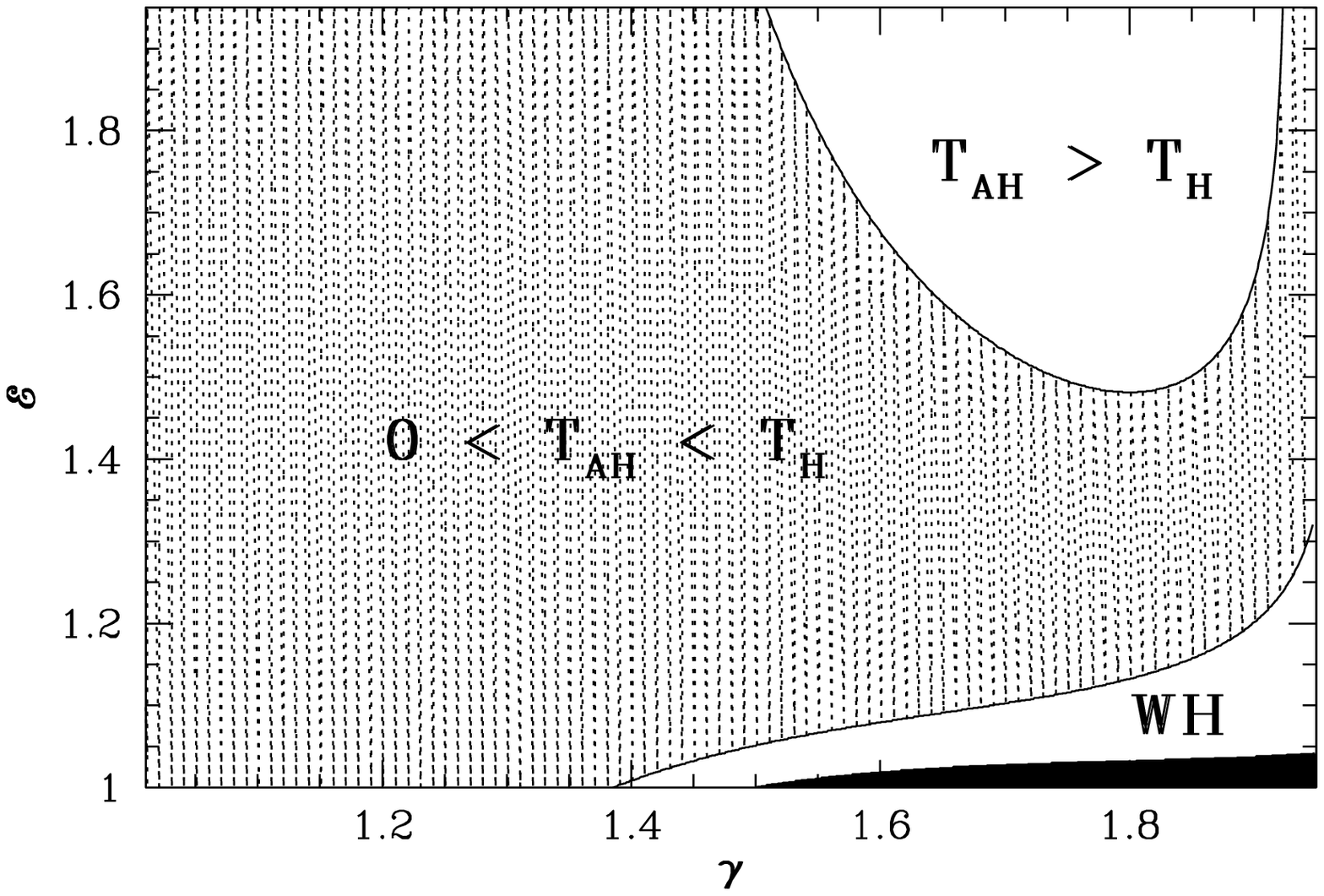,height=11cm,width=11cm,angle=0.0}}}
\centerline {{{\bf Figure 1:}
Parameter space classification for \egam for four different sets of values
of $T_{{AH}}$.}}
\end{figure}
\noindent
In the accompanying figure (taken from Das 2004a), 
the complete \egam  space is classified according
to the value of $\tau$. 
For accretion with \egam  taken from the lightly shaded
regions marked by $0<T_{{AH}}<T_H$,
the Hawking temperature dominates over its analogue counterpart.
For high $\gamma$, high ${\cal E}$ flow (`hot' accretion),
{\AHR} becomes the dominant process compared to the actual Hawking radiation
($T_{{AH}}>T_H$).
For low ${\cal E}$ and intermediate/high $\gamma$, acoustic white holes appear
(white region marked by {\bf WH}, where ${da_s/dr<du/dr}$ at $r_h$).
For \egam belonging to this region, one obtains outflow (outgoing solutions with Mach number
increasing with increase of $r$) only. This has also been verified by obtaining the
complete flow profile by integrating eq. (24).
The dark shaded
region (lower right corner of the figure) represents \egam for which $r_h$ comes out
to be physical ($r_h>1$) but
$\Phi_{123}^2$ becomes negative, hence $(du/dr)_h$ and $(da_s/dr)_h$ are not real and $T_{{AH}}$
becomes imaginary. Note that both $T_{{AH}}>T_H$ and white hole regions are obtained even for
$4/3<\gamma<5/3$, which is the range of values of $\gamma$ for most realistic flows of matter
around astrophysical black holes.
Hence the domination of
the analogue Hawking temperature over the actual Hawking temperature
and the emergence of
analogue white holes are real astrophysical phenomena.\\
\noindent
Note one {\it important} point that the accreting spherical black holes in general
relativity is the {\it only} analogue system found till date, where the analogue
temperature may be {\it higher} than the actual Hawking temperature. 
Newtonian or semi-Newtonian accreting system  does not show this behaviour.
Also note that both $T_H$ and $T_{AH}$ comes out to be
quite less compared to the macroscopic classical fluid temperature
of accreting matter
\footnote{The value of the flow temperature for non-radiative 
spherical accretion is found to be of the order of $10^{10}-10^{11}
{^oK}$.}. 
Hence black hole accretion system may not be a good
candidate to allow any observational test for detecting the analogue
radiation.
\subsection{Analogue effects in accretion disc}
\noindent
Calculation of $T_{AH}$ for disc accretion (as described in \S 9.2) is also possible.
One needs to calculate the acoustic metric and the Killing vectors for {\it axisymmetric}
space time. Then one needs to use those results to calculate the surface gravity and
analogue temperature. Initial attempts (Das \& Bilic, 2004, in preparation) shows that
although the geometry would be different (axisymmetric) than the spherical case, 
$T_{AH}$ will still be a function of $\left[u,a_s,du/dr,da_s/dr\right]$ at the 
sonic point, with different functional form as that of eq. (50). Hence for disc 
accretion also, the exact value of $T_{AH}$ can directly be calculated from the results
obtained in \S 9.2. However, as we discussed before, for axisymmetric accretion, more
than one sonic point is formed, hence the question now may arise that which sonic 
points are to be taken, to form the acoustic horizon. The answer seems to be quite 
straight forward. Firstly, all the middle sonic points would be excluded because no stable 
solutions pass through that. Secondly, as accretion can pass through the inner sonic 
point {\it only if} a shock forms (see \S 9.2), one should exclude the inner sonic 
points as well. This is because, shock production is dissipative and may violate the 
Lorenzian invariance. Hence, if the accretion is multi-transonic, which is the
situation for certain values of $\left[{\cal E},\lambda,\gamma,a\right]$,
{\it only} the flow passing 
through the {\it outer sonic point} will resemble an analogue system. For other values of  
 $\left[{\cal E},\lambda,\gamma,a\right]$, disc accretion is mono-transonic and hence
there will not be any confusion about the choice of the sonic points for producing 
the acoustic horizon.

However, although an observer at positive infinity will see only one acoustic 
black hole, there may be one alternative possibility for an observer situated
in between the two sonic points. An observer in between two sonic points 
(the inner sonic point and the outer sonic point) may perhaps see a white hole
and a black hole or two white holes (see Barcelo et. al. 2004 for discussion about 
somewhat similar situation). Detail study of this issue is beyond the scope of this article and will
be presented elsewhere.
\section{Acknowledgments}
This work is supported by the research grant 1P03D01827 from KBN. The author 
acknowledges useful discussions with Neven Bili$\acute{c}$, William Unruh,
Ted Jacobson, John Miller, Stefano Liberati, Paul Wiita and Paramita Barai.

\end{document}